\documentclass[pre,aps,twocolumn]{revtex4}

\usepackage{graphicx}
\usepackage{amsmath}
\usepackage{amssymb}
\usepackage{ifthen}
\usepackage{booktabs}

\usepackage{graphicx}
\usepackage{dcolumn}
\usepackage{bm}
\usepackage{longtable}

\newcommand{\bb}{\begin{equation}}
\newcommand{\ee}{\end{equation}}
\newcommand{\ba}{\begin{eqnarray*}}
\newcommand{\ea}{\end{eqnarray*}}

\begin{document}

\title{Microscopic determination of correlations in the fluid interfacial region in the presence of liquid-gas asymmetry}

\author{A.O.\ Parry}
\affiliation{Department of Mathematics, Imperial College London, London SW7 2BZ, UK}

\author{C.\ Rasc\'{o}n}
\affiliation{GISC, Departamento de Matem\'aticas, Universidad Carlos III de Madrid, 28911 Legan\'es, Madrid, Spain}
\affiliation{ICMAT, Campus Cantoblanco UAM, 28049 Madrid, Spain}

\begin{abstract}
In a recent article, we showed how the properties of the density-density correlation function and its integral, the local structure factor, in the fluid interfacial region, in systems with short-ranged forces, can be understood microscopically by considering the resonances of the local structure factor [Nat.Phys.~{\bf 15}, 287 (2019)]. Here we illustrate, using mean-field square-gradient theory and the more microscopic Sullivan density functional model, how this approach generalises when there is liquid-gas asymmetry, i.e. when the bulk correlation lengths of the coexisting liquid and gas phases are different. In particular, we are able to express the correlation function \textit{exactly} as a simple average of contributions arising from two effective Ising-symmetric systems referred to as the symmetric gas and symmetric liquid. When combined with our earlier results, this generates analytical approximations for the correlation function and the local structure factor, which are near indistinguishable from the numerical solution to the Ornstein-Zernike equations over the whole range of wave-vectors. Our results highlight how asymmetry affects the correlation function structure, and describes the crossover from a long-ranged Goldstone mode to short-ranged properties determined by the local density, as the wave-vector increases.
\end{abstract}

\maketitle

\section{Introduction}

Correlation functions play a vitally important role in characterising the equilibrium properties of the different possible phases of matter. At a very basic level, it is clear that the regular arrangement of atoms in a perfect solid lattice leads to long-ranged correlations in atomic positions. Simple liquids and gases display no long-ranged order, although there may be short-ranged order in a dense liquid, arising from local packing effects, which decays rapidly as the distance between the particles increases. Away from the immediate vicinity of the critical point, the pair correlation function in the liquid and gas phases is isotropic and decays exponentially on a microscopic scale set by the appropriate bulk correlation length. There are two well known scenarios in which this simple picture is complicated. Complex fluids, such as liquid crystals, display a much wider variety of phases which are distinguished by the orientational as well as positional order. Secondly, even for simple fluids, the presence of an interface separating coexisting liquid and gas phases drastically alters the nature of atomic correlations. This arises directly from the thermal fluctuations of the interface, controlled by the surface tension $\sigma$, which lead to much longer-ranged correlations in the vicinity of the interface \cite{BLS1965,Wertheim1976,Weeks1977,Evans1979,Rowlinson1982,Aarts2004}. If the interface is pinned by a gravitational field, correlations along the interface decay on a scale set by the capillary length, which is about a $mm$ for molecular fluids. Indeed, in the absence of pinning due to gravity or a nearby wall (say), translations of a free interface cost no energy, resulting in a scale-free Goldstone mode. In this case, exact microscopic sum-rules determine that the two-dimensional Fourier transform of the density-density (pair) correlation function $G(z,z';q)$, where $q$ is the wave-vector parallel to the interface and $z$ the co-ordinate normal to the interface, must diverge, when $q\to 0$, as \cite{Wertheim1976,Weeks1977}
\begin{equation}
G(z,z';q)\;\approx\; \frac{\,\rho'(z)\,\rho'(z')\,}{\beta\sigma q^2}
\label{GM}
\end{equation}
Here, $\rho(z)$ denotes the density profile, $\rho'(z)=d \rho(z)/dz$ its derivative and $\beta=1/k_B T$. This result is of profound significance to the fundamental statistical mechanics of fluids at interfaces. In particular, it tallies precisely with the mesoscopic capillary-wave theory of the interface, which models it simply as a structureless surface under tension -- a description which is extremely accurate at length-scales much larger than the underlying microscopic bulk correlation length \cite{BLS1965,Weeks1977}.

The question of how correlations near a fluid interface behave at larger values of the wave-vector $q$, that is, how the Goldstone-mode divergence (\ref{GM}) is modified, has received considerable attention over the last few decades \cite{Rochin1991,Napiorkowski1993,Parry1994,Robledo1997,Mecke1999,Fradin2000,Blokhuis2008,
Blokhuis2009,Parry2014,Hofling2015,Parry2015,Chacon2016,Parry2016,Hernandez2018,Parry2019}. In recent articles, we have shown how one may use microscopic Density Functional Theory (DFT) to determine the properties of $G(z,z',q)$ and its integral $S(z;q)$ (the local structure factor) over the whole range of wave-vectors for systems with short-ranged forces \cite{Parry2019,Parry2019b}. These new insights arise from noting that the properties of the local structure factor and correlation function are strongly constrained due to the presence of resonances in $S(z;q)$, which occur at specific values of the wave-vector. This approach allows us to understand analytically the crossover from the Goldstone mode at small $q$ to bulk-like behaviour at larger $q$, without having to introduce mesoscopic ideas such as a wave-vector dependent surface tension. In our previous work, we focussed largely on model systems which display a simple Ising symmetry. In the present paper, we show how this analysis can be extended to the more realistic case where there is liquid-gas asymmetry, i.e. where the correlation lengths of the coexisting bulk liquid and gas phases are different. In particular, using simple mean-field square-gradient theory we show that there is an exact construction which determines the correlation function and local structure factor at the interface (specifically, where the gradient $\rho'(z)$ is largest) in terms of averages of the corresponding quantities of an effective Ising symmetric gas and an effective Ising symmetric liquid. This allows us to determine analytically the properties of $G(0,0,q)$ and $S(0,q)$ over the the whole range of wave vectors. We show that same approach also applies to the more microscopic Sullivan density functional model \cite{VanKampen1964,Percus1964,Sullivan1979,Sullivan1981} allowing us to determine the form of the correlation function and local structure factor using the accurate Carnahan-Starling equation of state for hard-spheres.

Our paper is arranged as follows. We begin with square-gradient theory and recap the main results for $G$ and $S$ which apply to systems with a perfect Ising symmetry. We then allow for asymmetry between the bulk liquid and gas phases, and describe a construction which allows one to determine exactly the correlation function at the interface as an average of two equally weighted Ising symmetric systems -- one corresponding to a gas and one to a liquid. When combined with the earlier results for symmetric systems, this leads to an analytical approximation for $G(0,0;q)$, which we test against the numerical solution of the Ornstein-Zernike (OZ) equation for a model square-gradient potential with a wide range of asymmetries. A simple extension of the analysis generates an expression for the local structure factor $S(0;q)$ at the interface, which we compare against the numerical solution of the OZ equation. We repeat this exercise for the Sullivan model showing how the analytical approximations for $G(0,0;q)$ and $S(0;q)$ capture very accurately the properties of the correlation function and local structure factor over the whole range of wave-vectors. Finally, the implications and physical interpretation of our results are discussed.

\section{Correlations in the presence of liquid-gas asymmetry}
 
\subsection{Density Functional Theory formalism} 

Within the framework of DFT, the equilibrium density profile and correlation functions of an inhomogeneous fluid can be determined from a Grand Potential functional \cite{Evans1979}
\begin{equation}
\Omega[\rho]=F[\rho]-\int\! d{\bf r}\;\,\big(\mu-V_{ext}({\bf r})\big)\,\rho({\bf r})
\end{equation}
where $\mu$ is the chemical potential, $V_{ext}({\bf r})$ is the external field, and $\rho({\bf r})$ is the density distribution. All the information concerning fluid-fluid forces is contained within the Hemlholtz free-energy functional $F[\rho]$ for which various mean-field approximations can be made. The equilibrium density profile is obtained from minimization of the Grand Potential
\begin{equation}
\frac{\delta\Omega[\rho]}{\delta\rho}\;=\;0
\end{equation}
and the direct correlation function is obtained as the second functional derivative
\begin{equation}
C({\bf r},{\bf r}')\;=\;\frac{1}{k_B T}\,\frac{\delta^2F[\rho]}{\delta \rho({\bf r})\,\delta\rho({\bf r}')}
\end{equation}
which must be evaluated at the equilibrium fluid density. Hereafter, we set $k_B T=1$. From the direct correlation function, we can obtain the density-density correlation function $G({\bf r},{\bf r}')$ from solution of the inhomogeneous OZ equation
\begin{equation}
\int\! d{\bf r}'' \;C({\bf r},{\bf r}'')\,G({\bf r}'',{\bf r}')\;=\;\delta({\bf r}-{\bf r}')
\end{equation}
We consider a free planar interface (i.e., under a vanishing external field) separating coexisting liquid and gas phases, for which the equilibrium density profile $\rho(z)$ is a function only of the distance $z$ normal to the interface. Translational invariance along the interface means that both the direct and density-density correlation depend on the co-ordinates $z$ and $z'$ of the two particles and the parallel separation between them. It is then convenient to consider the 2D Fourier transforms $C(z,z';q)$ and $G(z,z';q)$, where $q$ is the modulus of the wave-vector parallel to the interface. In this case, the OZ equation reduces to 
\begin{equation}
\int\! d z'' \; C(z,z''; q)\,G(z'',z'; q)\;=\;\delta(z-z')
\end{equation}
which we shall seek to solve for different model density functionals. The integral of the density-density correlation function defines the local structure factor
\begin{equation}\label{S}
S(z;q)\;= \int\! d z\;\,G(z,z'; q)
\end{equation}
whose $q=0$ limit identifies the local compressibility $S(z;0)=\partial\rho(z)/\partial\mu$. Therefore, the local structure factor may also be determined directly by solving the integrated OZ equation
\begin{equation}
\int\! d z' \; C(z,z'; q)\,S(z'; q)\;=\;1
\end{equation}
Finally, we mention that the surface tension for the free interface is determined via the thermodynamic definition
\begin{equation}\label{sig}
\sigma\;=\;\frac{\Omega+pV}{A}
\end{equation}
where $\Omega$ is the equilibrium value of the Grand Potential, $p$ is the pressure, and $V$ and $A$ are the volume and interfacial area, respectively.

In this paper, we consider two simple models of the interfacial region applicable to systems with short-ranged forces: Square-gradient theory and the more microscopic Sullivan local density functional \cite{Sullivan1979,Sullivan1981}. Neither model accounts for packing effects at the molecular scale, although such effects are relatively minor for a free liquid-gas interface. This is particularly the case if we restrict attention to temperatures above the intersection of the Fisher-Widom line and the liquid-gas coexistence curve, where the density profile decays monotonically on the liquid side \cite{Evans1993}. Being mean-field in nature, neither model accounts for the capillary-wave induced broadening of the liquid-gas interface. In three dimensions, however, this broadening is extremely weak since the interfacial width is anticipated to increases as $\sqrt{\ln A}$ \cite{BLS1965}. Consequently, the density profiles predicted by both models will be at least qualitatively similar to those seen in simulation studies of systems of truncated Lennard-Jones forces (say). The merits of both these simple models is that they account for both bulk and interfacial behaviour, and therefore allow us to consider the modification to the Goldstone mode divergence (\ref{GM}) as the wave-vector $q$ is increased to the scale of the inverse bulk correlation length, where the standard capillary-wave picture of the interface breaks down. Here, as mentioned earlier, we shall concentrate on allowing for liquid-gas asymmetry, as is pertinent to real fluids.

\subsection{Square-gradient model}

The first model we consider is a mean-field square-gradient theory based on the Grand Potential functional \cite{Evans1979,Rowlinson1982}
\begin{equation}
\Omega[\rho]\;=\int\!{d{\bf r}}\;\left(\frac{f}{2}(\nabla\rho)^2+\Delta\phi(\rho)\right)
\end{equation}
where the coefficient $f$ of the gradient term is hereafter set to unity, since it does not appear in our final results. Below a bulk critical temperature $T_c$, the  bulk free-energy density $\phi(\rho)$ has a double-well structure modelling the coexistence of bulk liquid and gas phases with densities $\rho_l$ and $\rho_g$. Thus $\phi'(\rho_g)=\phi'(\rho_l)=0$, with coexistence demanding that $\phi(\rho_g)=\phi(\rho_l)$. The shifted potential $\Delta\phi(\rho)=\phi(\rho)-\phi(\rho_l)$ conveniently subtracts the bulk contribution. In general, the potential will also have a maximum at some intermediate density $\rho_0$, where $\Delta\phi'(\rho_0)=0$. For Ising symmetric potentials, this is obviously equivalent to the mid-point or critical density $\rho_0=(\rho_g+\rho_l)/2$. The curvatures of the potential $\phi''(\rho_b)=\kappa_b^2$ determine the inverse correlation lengths $\kappa_b=1/\xi_b$ of the bulk liquid ($b=l$) and bulk gas ($b=g$) phases. These determine the exponential decay of the bulk pair correlation function $G_b(r)\propto exp (-\kappa_b r)/r$, where $r$ is the distance between the particles. The three dimensional Fourier transform of $G_b(r)$ defines the two bulk structure factors, which in square-gradient theory have a simple Lorentzian form
\begin{equation}
S_b(q)\;=\; \frac{S_b(0)}{1+\xi_b^2q^2}
\end{equation}
where $S_b(0)=1/\Delta \phi''(\rho_b)$ identifies the bulk compressibility $S_b(0)=\partial\rho/\partial\mu$. It is also convenient to consider the 2D (Hankel) Fourier transform when the particles sit in the {\it same} $z$ plane 
\begin{equation}
G_b(q)\;=\;\frac{1}{2\sqrt{\kappa_b^2+q^2}}
\end{equation}
which was referred to as $G_b(0;q)$ in \cite{Parry2019}.
We now consider that a planar interface of macroscopic area separates the coexisting liquid and gas phases, and is located near the $z=0$ plane. At mean-field level, no additional pinning field needs to be specified. The equilibrium density profile follows from solution of the Euler-Lagrange equation
\begin{equation}
\rho''(z)\;=\; \Delta\phi'(\rho)
\end{equation}
where the prime denotes differentiation with respect to the argument shown. This is solved subject to boundary conditions $\rho(-\infty)=\rho_g$ and $\rho(\infty)=\rho_l$ with the origin chosen to correspond to the maximum of the density gradient, i.e. $\rho''(0)=0$. Thus, the density at the origin corresponds to $\rho_0$, the maximum in the potential $\Delta\phi(\rho)$. This is not a definition of an interfacial collective coordinate; rather, a convenient choice for the origin of the co-ordinates. The Euler-Lagrange equation has a first integral
 \begin{equation}
\rho'(z)\;=\; \sqrt{ 2\Delta\phi(\rho)}
\end{equation}
and substitution into $\Omega[\rho]$ determines the surface tension via (\ref{sig}) as
 \begin{equation}
\sigma\;=\; \int_{-\infty}^{\infty}\!\!\!dz\;\,\rho'(z)^2 
\end{equation}
or, equivalently,
\begin{equation}
\sigma\;=\; \int_{\rho_g}^{\rho_l}\!\!d\rho\; \sqrt{ 2\Delta\phi(\rho)}
\end{equation}
The direct correlation function of the inhomogeneous fluid is given by the delta function operator
\begin{equation}
C({\bf r},{\bf r}')\;=\;\big(-\nabla^2_{\bf {r}}+\Delta\phi''(\rho(z))\big)\;\delta({\bf r}-{\bf r}')
\end{equation}
which has the 2D Fourier transform
\begin{equation}
C(z,z';q)\;=\;\big(-\partial^2_z+q^2+\Delta\phi''(\rho(z))\big)\;\delta(z-z')
\end{equation}
Hence, the OZ equations for the correlation function and local structure factor reduce to
\begin{equation}\label{OZG}
\big(-\partial^2_z+q^2+\Delta\phi''(\rho(z))\big)\;G(z,z';q)\;=\;\delta(z-z')
\end{equation}
and 
\begin{equation}\label{OZS}
\big(-\partial^2_z+q^2+\Delta\phi''(\rho(z))\big)\;S(z;q)\;=\;1
\end{equation}

\subsection{Correlations with Ising symmetry}

It has long been known that, for all potentials $\phi(\rho)$, the correlation function obtained from the OZ equation displays the anticipated Goldstone mode divergence 
(\ref{GM}) as $q\to 0$ \cite{Evans1981}. Hence, the local structure factor also diverges as $S(z;q)\approx\Delta\rho\,\rho'(z)/\sigma q^2$ in the same small wave-vector limit. Here, $\Delta\rho=\rho_l-\rho_g$ is the difference between the bulk densities. Recently, however, new insights into the full wave-vector behaviour of the correlation function and local structure factor have emerged from recognising that the local structure factor has, in addition to the Goldstone mode, a hierarchy of resonances occurring at $q\xi_b=\sqrt{3},\sqrt{8}\sqrt{15}\cdots$, on the liquid ($b=l$) and gas ($b=g$) sides \cite{Parry2019}. These strongly constrain the allowed behaviours of both $G$ and $S$. Let us begin with the case of perfect Ising (lattice-gas) symmetry, for which the bulk correlation functions and hence $\xi_b$, $S_b(q)$ and $G_b(q)$ are identical in the liquid and gas phases. In this case, the existence of the resonances means that, in addition to (\ref{S}), there is another relation between the local structure and correlation function which may be expressed as a sum over the resonances. Specifically, for all potentials $\phi(\rho)$ which have an analytic expansion about the bulk density, the local structure factor can be written
\begin{equation}\label{Res}
\begin{split}
S(z;q)\;=\;S_b(q)\,+\frac{\Delta\rho\rho'(z)}{\sigmạ_1q^2(1+\xi_b^2 q^2)}\;+\\[.25cm] \frac{\Delta\rho}{\rho'(0)}\,\sum_{n=2}^\infty\,\frac{\sigma}{\sigma_n}\,\frac{G(0,z;q)-G\left(0,z;\sqrt{n^2-1}\kappa_b\right)}{(1+\xi_b^2 q^2)(1-\frac{\xi_b^2 q^2}{n^2-1})}
\end{split}
\end{equation}
where, again, we emphasise that the origin corresponds to the maximum in $\rho'(z)$. The resonances are weighted by generalised surface tension-like coefficients $\sigma_n$, the values of which are determined by the correlation function, and satisfy the summation condition $1/\sigma_1+1/\sigma_2+\dots= 1/\sigma$. This automatically ensures that the local structure factor has the required Goldstone mode divergence at small $q$. From this relation, two extremely accurate approximations for the wave-vector dependence of the  correlation function and local structure factor at the origin emerge:
\begin{equation}
G(0,0,q)\;\approx\; G_b(q)\;+\;\frac{\rho'(0)^2}{\sigma q^2}\,\frac{G_b(q)}{G_b(0)}
\label{Gapprox}
\end{equation}
and 
\begin{equation}
S(0;q)\;\approx\; S_b(q)\;+\;\frac{\Delta\rho\rho'(0)}{\sigma q^2}\,\frac{S_b(q)}{S_b(0)}
\label{Sapprox}
\end{equation}
The expression for $G(0,0;q)$ implies that the real-space decay of the correlation function along the interface may be regarded as the bulk decay plus a convolution between a bulk-like term and the Goldstone mode, arising from interfacial fluctuations. 
Recall that the results (\ref{Gapprox}) and (\ref{Sapprox}) are exact for the standard Landau quartic potential \cite{Parry2019}. The expression for $G(0,0;q)$ is also exact for the trigonometric potential $\Delta\phi(\rho)\propto \sin^2\left(\pi\,(\rho-\rho_g)/\Delta\rho\right)$. In general, these approximations always capture correctly the low and high $q$ limits, and are typically never more than a few percent away from the result obtained from numerical solution of the OZ equations (\ref{OZG}) and (\ref{OZS}). This is particularly true for the expression for $G(0,0;q)$ which, for example, has a maximum error of less than $0.5\%$ for the potential $\phi(\rho)$ modelling an interface near a tricritical point \cite{Parry2019b}. For practical purposes, they can be considered the full analytical solutions for the correlation function and local structure factor at the origin, where these functions take their maximum values in the interfacial region.

\subsection{Correlations with liquid-gas asymmetry}

\begin{figure*}[t]
\includegraphics[width=\columnwidth]{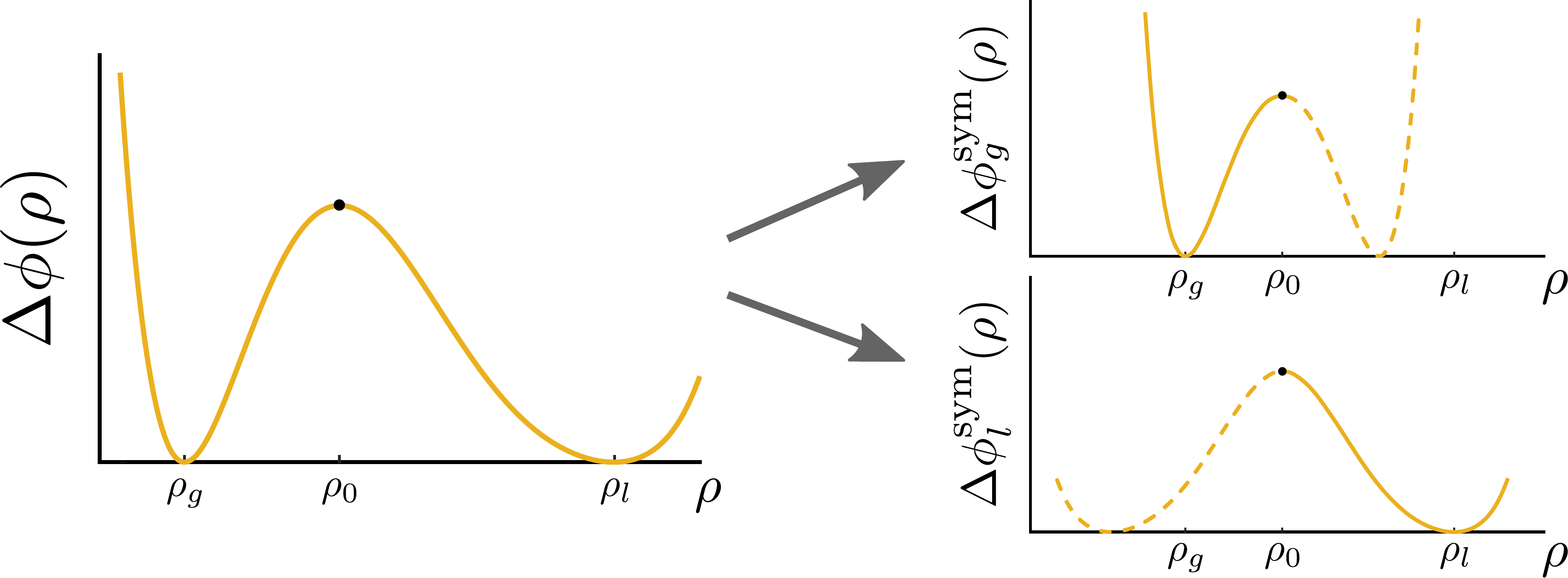}
\caption{\label{Fig1} An asymmetric potential $\Delta\phi(\rho)$ and two symmetric gas and liquid potentials $\Delta\phi_g^\textit{sym}(\rho)$ and $\Delta\phi_l^\textit{sym}(\rho)$ constructed from it by reflecting $\Delta\phi(\rho)$ about the density $\rho_0$, where it is a maximum (black dot).
}
\end{figure*}

We now turn attention to systems with liquid-gas asymmetry. That is, systems for which the correlation lengths $\xi_b$, and hence $G_b(q)$ and $S_b(q)$, are different in the bulk liquid and gas phases. A potential $\Delta\phi(\rho)$ modelling this is shown schematically in Fig.~\ref{Fig1} and has a larger curvature at the bulk gas density, $\Delta\phi''(\rho_g)> \Delta\phi''(\rho_l)$, so that $\xi_l>\xi_g$ as pertinent to real fluids. Also, the density $\rho_0$ at which $\Delta\phi(\rho)$ has a maximum is closer to the bulk gas density than to the liquid. We now describe a simple construction which allows us to determine the correlation function $G(0,0;q)$ in this asymmetric system, where again the origin is chosen to be the position where $\rho(0)=\rho_0$ and the density gradient is a maximum. First, we note the surface tension can be divided trivially into contributions from the "gas" and "liquid" regions either side of the origin:
\begin{equation}
\sigma\;=\; \int_{\rho_g}^{\rho_0}\!\!\!d\rho\; \sqrt{ 2\Delta\phi(\rho)}\,+\int_{\rho_0}^{\rho_l}\!\!\!d\rho\; \sqrt{ 2\Delta\phi(\rho)}
\end{equation}
Next, from the asymmetric potential $\Delta\phi(\rho)$, we construct two effective Ising symmetric potentials as shown in Fig.~\ref{Fig1}: A {\it symmetric gas} potential $\Delta\phi_g^\textit{sym}(\rho)$, which is defined to be the same as $\Delta\phi(\rho)$ for $\rho<\rho_0$ together with its mirror reflection for $\rho>\rho_0$, and a {\it symmetric liquid} potential $\Delta\phi_l^\textit{sym}(\rho)$, defined to be the same as $\Delta\phi(\rho)$ for $\rho>\rho_0$ together with its mirror reflection for $\rho<\rho_0$. We can define and determine the physical properties for each of these symmetric systems. For example, the surface tensions of the symmetric gas and symmetric liquid potentials are clearly given by 
\begin{equation}
\sigma_g^\textit{sym}\;=\; 2\int_{\rho_g}^{\rho_0}\!\!\!d\rho\;\, \sqrt{ 2\Delta\phi(\rho)}
\end{equation}
and 
\begin{equation}
\sigma_l^\textit{sym}\;=\; 2\int_{\rho_0}^{\rho_l}\!\!\!d\rho\;\, \sqrt{ 2\Delta\phi(\rho)}
\end{equation}
respectively. Thus, the true surface tension $\sigma$ can be regarded as the \textit{average} of the two Ising symmetric tensions:
\begin{equation}
\sigma\;=\;\frac{1}{2}\,\big(\sigma_g^\textit{sym}+\sigma_l^\textit{sym}\big)
\label{sigmaasym}
\end{equation}
Similarly, we can determine the correlation functions from solution of  
\begin{equation}
\Big(-\partial^2_z+q^2+\Delta\phi_b^\textit{sym }{''}\big(\rho(z)\big)\Big)\,G_b^\textit{sym}(z,z';q)=\delta(z-z')
\end{equation}
for the symmetric gas ($b=g$) and symmetric liquid ($b=l$) phases. Here, $\rho(z)$ is the profile for the corresponding symmetric potential, which are simply related to the true density profile.
It is then easy to see from simple matching of the solutions for $z>0$ and $z<0$ that the (inverse of the) correlation function of the asymmetric potential, at the origin, can be written {\it exactly} as the average 
\begin{equation}
\frac{1}{G(0,0;q)}\;=\;\frac{1}{2}\left(\frac{1}{G_g^\textit{sym}(0,0;q)}+\frac{1}{G_l^\textit{sym}(0,0;q)}\right)
\label{Gasym1}
\end{equation}
which can also be written as
\begin{equation}
G(0,0;q)\;=\; \textup{f}_g^*(q)\;G_g^\textit{sym}(0,0;q)+\textup{f}_l^*(q)\;G_l^\textit{sym}(0,0;q)
\label{Gasym1new}
\end{equation}
where 
\begin{equation}\label{fstar}
\textup{f}_g^*(q)\;=\;\frac{G_l^\textit{sym}(0,0;q)}{G_g^\textit{sym}(0,0;q)+G_l^\textit{sym}(0,0;q)}
\end{equation}
and $\,\textup{f}_g^*(q)+\textup{f}_l^*(q)=1$. The utility of the exact result (\ref{Gasym1}) is that we can combine it with the extremely accurate analytical approximations 
\begin{equation}
G_b^\textit{sym}(0,0;q)\approx G_b(q)+\frac{\rho'(0)^2}{\sigma_b^\textit{sym} q^2}\frac{G_b(q)}{G_b(0)}
\label{Gasym2}
\end{equation}
which, we stress, only involves properties defined using the true asymmetric potential and density profile. This leads to 
\begin{equation}\label{Gapprox2}
\frac{1}{G(0,0;q)}\;\approx\;\frac{1}{2}\,\sum_{b=g,l}\,\frac{1}{G_b(q)\left(1+
\frac{\rho'(0)^2}{\sigma_b^\textit{sym}\,q^2\,G_b(0)}\right)}
\end{equation} 

This approximation is one of the main new results of our paper and, as we shall see, is effectively the analytical solution for the correlation function for a potential with arbitrary liquid-gas asymmetry. Indeed, it is exact, for example, when the double-well potential is made from matching two Landau quartic potentials with different liquid and gas correlation lengths, or by matching a Landau quartic potential for one well and a trigonometric potential for the other.\\

\begin{figure*}
\includegraphics[width=\columnwidth]{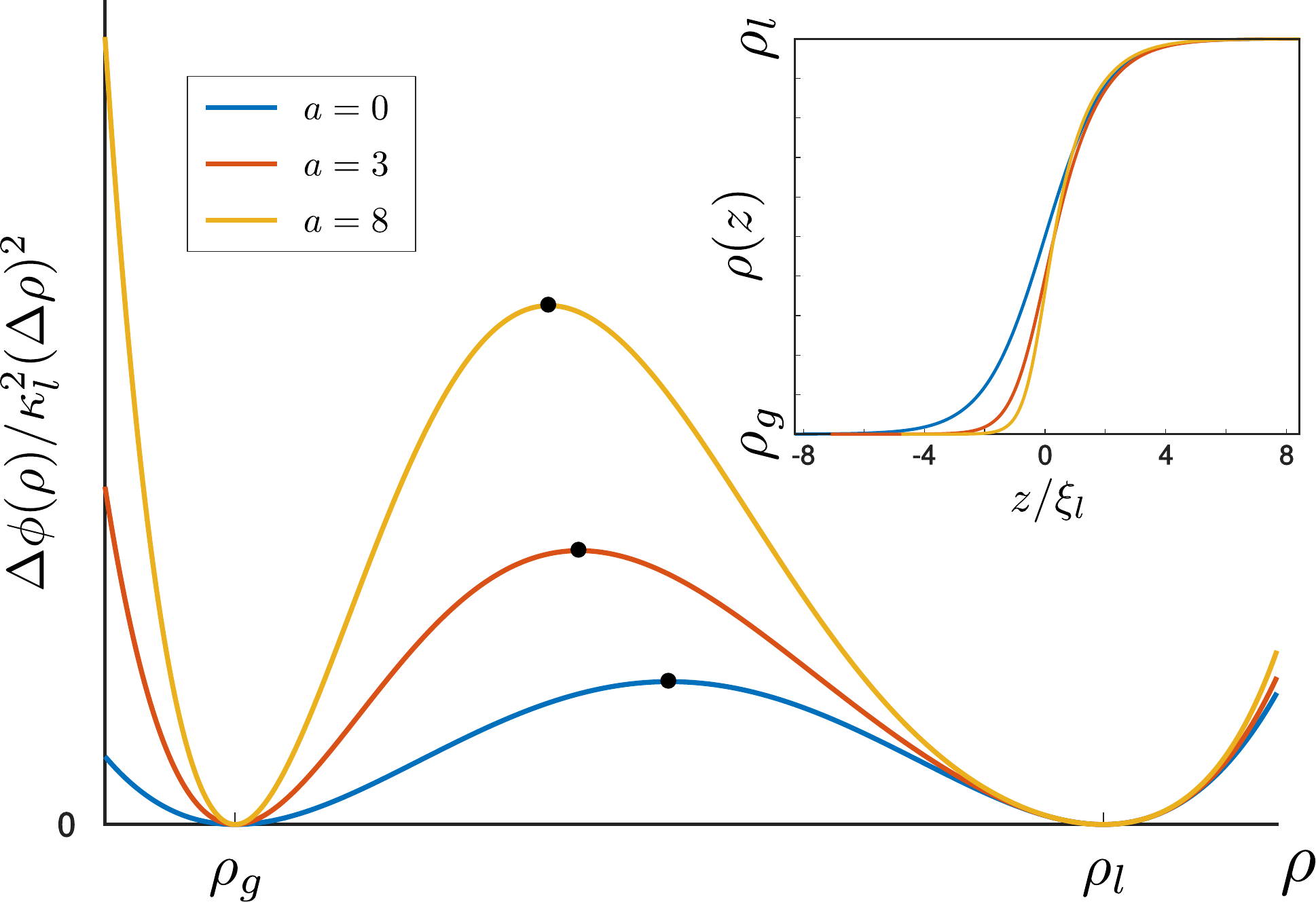}
\caption{\label{Fig2} Asymmetric model potential (\ref{phiasym}) for three values of the parameter $a$. The value $a=0$ corresponds to Ising symmetry ($\xi_g=\xi_l$), while $a=3$ and $a=8$ correspond to strong asymmetry: $\xi_l=2\,\xi_g$ and $\xi_l=3\,\xi_g$, respectively.
The black dots represent the location of the maximum, which corresponds to the origin of co-ordinates for the density profiles $\rho(z)$, shown in the inset.
}
\end{figure*}

\begin{figure*}
\includegraphics[width=\columnwidth]{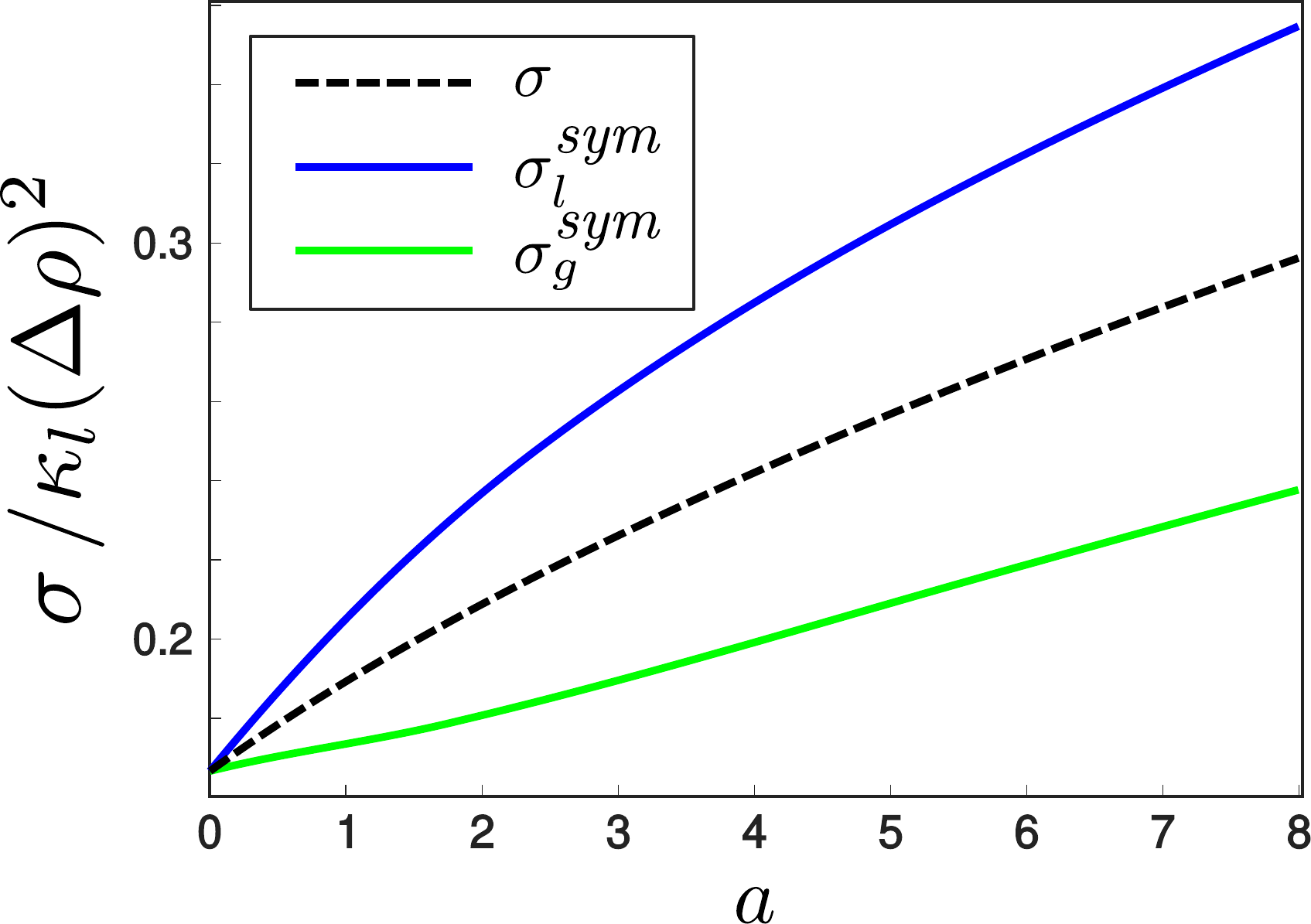}
\caption{\label{Fig3} Surface tensions $\sigma_g^\textit{sym}$ and $\sigma_l^\textit{sym}$ for the symmetric gas and liquid as a function of the asymmetry parameter $a$ for the potential (\ref{phiasym}). The true tension $\sigma$ is the average (dashed line).
}
\end{figure*}

\begin{figure*}
\includegraphics[width=\columnwidth]{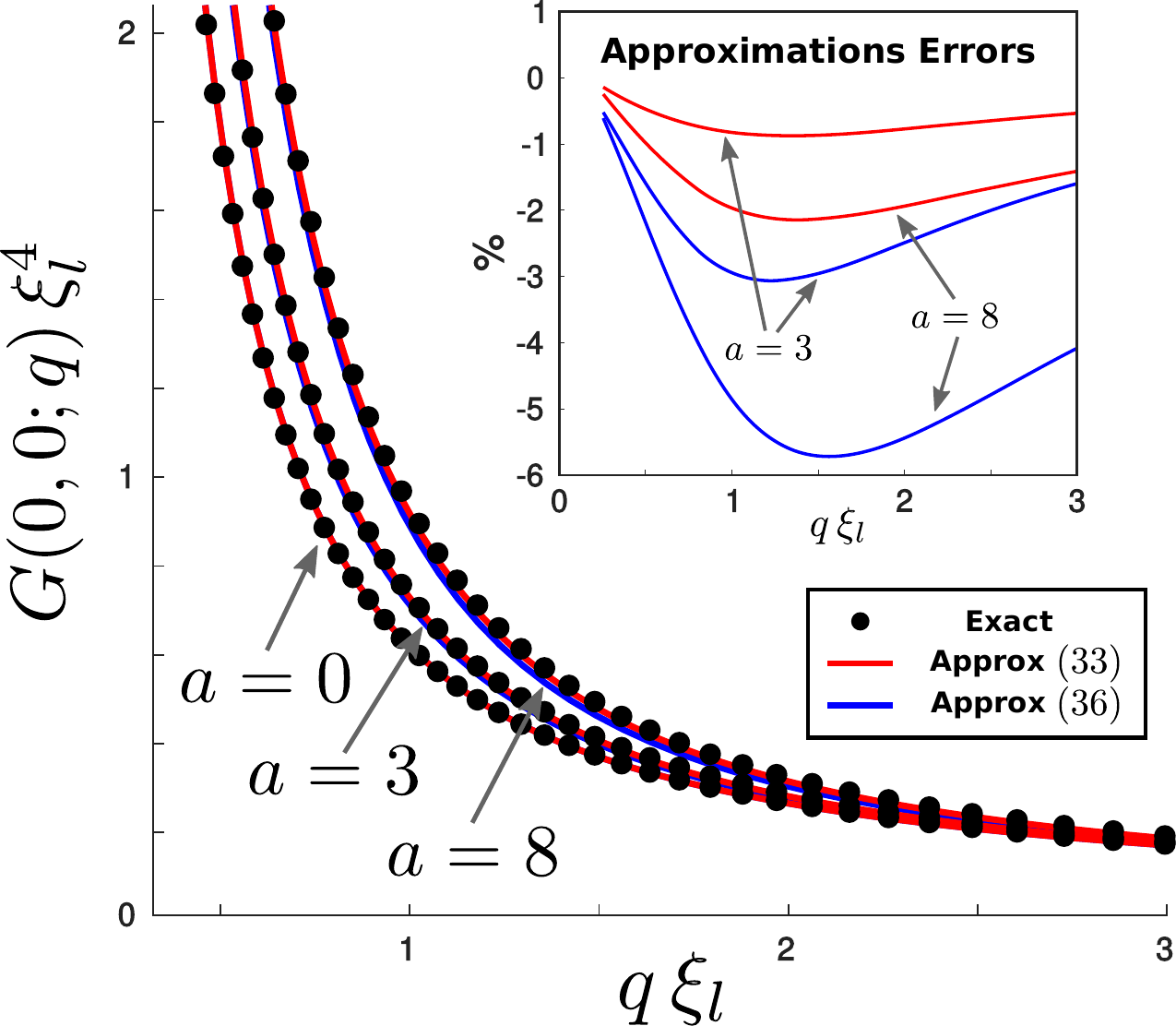}
\caption{\label{Fig4} Comparison of the exact numerical solution (dots) of the OZ equation (\ref{OZG}) for $G(0,0;q)$ using the asymmetric potential (\ref{phiasym}) with the analytical approximation (\ref{Gapprox2}) for three different values of the asymmetry parameter $a$ (red lines). The even simpler approximation (\ref{Gapprox3}), which uses $\sigma_b^\textit{sym}\approx \sigma$, is also shown (blue lines). Percentage errors are shown in the inset. Both approximations are exact for the Ising symmetric case $a=0$.
}
\end{figure*}

To illustrate the accuracy of the above analytical approximation, we apply it to the model six-order polynomial potential (see Fig.~\ref{Fig2}) 
\begin{equation}
\Delta\phi(\rho)\;=\;\frac{\kappa_l^2(\rho-\rho_g)^2(\rho-\rho_l)^2}{2(\Delta\rho)^2}\left(1+a\,\frac{(\rho-\rho_l)^2}{(\Delta\rho)^2}\right)
\label{phiasym}
\end{equation}
which contains a dimensionless parameter $a$ controlling the asymmetry between the bulk correlation lengths:
\begin{equation}
\xi_l=\sqrt{1+a\,}\;\xi_g
\end{equation} 
When $a=0$, the potential reduces to the standard Landau quartic potential, for which (\ref{Gapprox}) and (\ref{Sapprox}) are exact. When $a>0$, the model is asymmetric and
no longer exactly solvable. For example, for $a=3$, we have $\xi_l=2\,\xi_g$, while for $a=8$, we have $\xi_l=3\,\xi_g$. Corresponding profiles are shown in the inset of Fig.~\ref{Fig2}. The values of the symmetric liquid and symmetric gas surface tensions are shown in Fig.~\ref{Fig3} as a function of the asymmetry parameter. Fig.~\ref{Fig4} then compares the analytical approximation (\ref{Gapprox2}) with the full numerical solution of the OZ equation (\ref{OZG}), demonstrating its extraordinary accuracy over the full range of wave-vectors. For completion, we also show the comparison obtained when we approximate both symmetric surface tensions by the equilibrium value, $\sigma_b^\textit{sym}\approx \sigma$, in (\ref{Gapprox2}). That is, we make the further approximation
\begin{equation}\label{Gapprox3}
\frac{1}{G(0,0;q)}\;\approx\;\frac{1}{2}\,\sum_{b=g,l}\,\frac{1}{G_b(q)\left(1+
\frac{\rho'(0)^2}{\sigma\,q^2\,G_b(0)}\right)}
\end{equation} 
which does not require the evaluation of the symmetric surface tensions, and can be easily implemented in simulation studies. Although this latter approximation is about three times less accurate than the original one, it is still a remarkable good description of the full wave-vector dependence, encompassing the correct small and large-$q$ behaviour.

\subsection{The local structure factor with liquid-gas asymmetry}

\begin{figure*}
\includegraphics[width=\columnwidth]{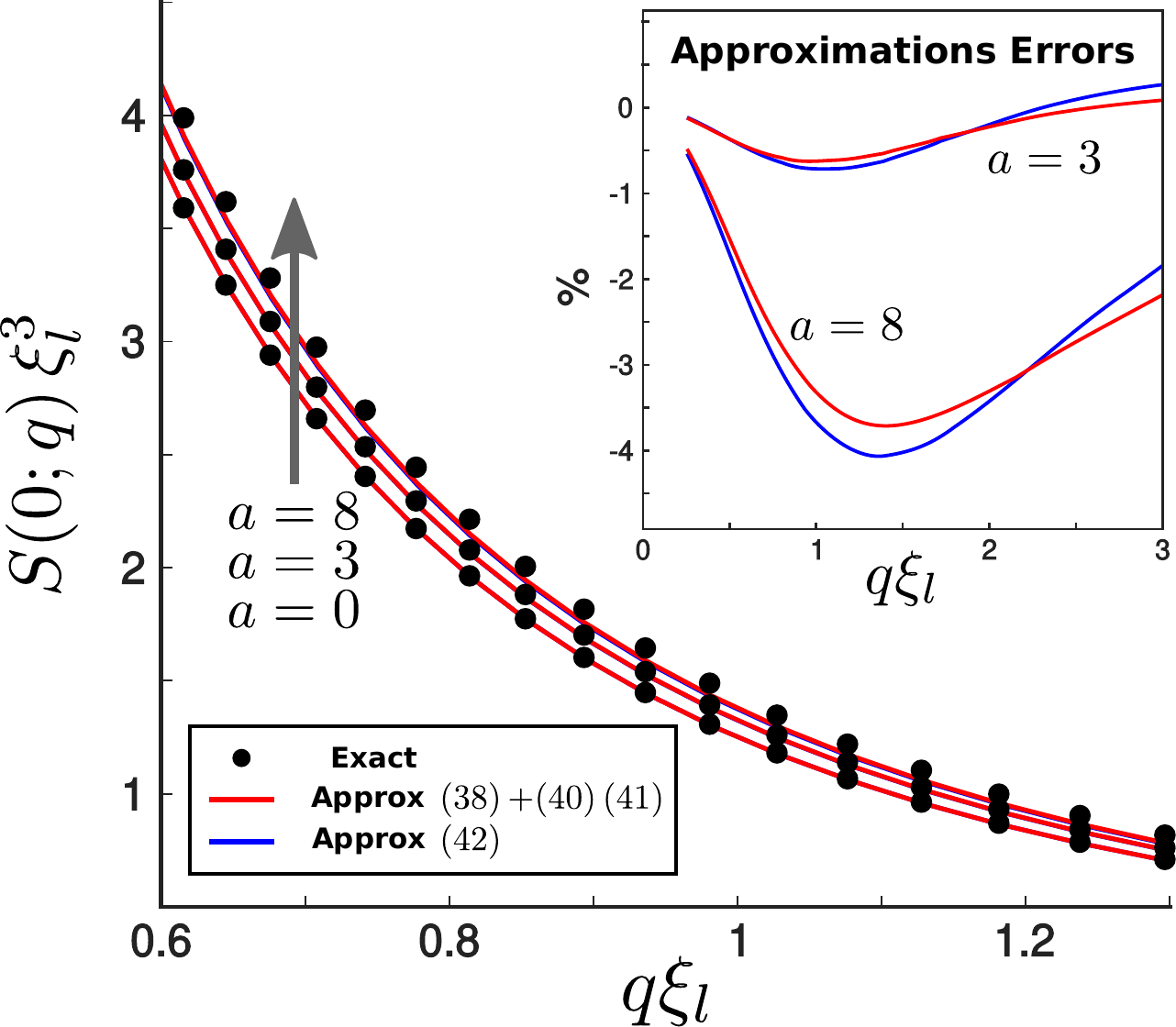}
\caption{\label{Fig5}  Comparison of the exact numerical solution (dots) of the OZ equation (\ref{OZS}) for $S(0;q)$ using the asymmetric potential (\ref{phiasym}) with the analytical approximation based on (\ref{Sasym}) together with (\ref{Ssymg}) and (\ref{Ssyml}), for three different values of the asymmetry parameter $a$ (red lines). The alternative approximation (\ref{Snat}) is also shown (blue lines) with almost identical results. Percentage errors are shown in the inset. Both approximations are exact for the Ising symmetric case $a=0$.
}
\end{figure*}

The above analysis extends to the determination of the local structure factor. Either side of the origin, the correlation function decays as 
\begin{equation}\label{Gdecay}
G(0,z;q)\;=\;\frac{G(0,0;q)}{G_b^\textit{sym}(0,0;q)}\;\,G_b^\textit{sym}(0,z;q)
\end{equation}
for $z>0$ ($b=l$) and $z<0$ ($b=g$). Integration of (\ref{Gdecay}) determines that the local structure at the origin can be written {\it exactly} as the weighted sum of contributions from the symmetric gas and symmetric liquid:
\begin{equation}\label{Sasym}
S(0;q)\;=\;\textup{f}_g^*(q)\;\,S_g^\textit{sym}(0;q)
+\;\textup{f}_l^*(q)\;\,S_l^\textit{sym}(0;q)
\end{equation}
which, pleasingly, has the same weightings as the result (\ref{Gasym1new}) for the correlation function.
This is an exact result if we determine the local structure factors of the symmetric gas and liquid from solution of 
\begin{equation}
\big(-\partial^2_z+q^2+\Delta\phi_b^\textit{sym }{''}(\rho(z))\big)\;S_b^\textit{sym}(z,z';q)\;=\;1
\end{equation}
However, as with our discussion of the correlation function, there is no need to do this. Instead, we may use in (\ref{Sasym}) the very accurate approximations
\begin{equation}
S_g^\textit{sym}(0;q)\;\approx\; S_g(q)+\frac{2(\rho_0-\rho_g)\rho'(0)}{\sigma_g^\textit{sym} q^2}\;\frac{S_g(q)}{S_g(0)}
\label{Ssymg}
\end{equation}
and 
\begin{equation}
S_l^\textit{sym}(0;q)\;\approx\; S_l(q)+\frac{2(\rho_l-\rho_0)\rho'(0)}{\sigma_l^\textit{sym} q^2}\;\frac{S_l(q)}{S_l(0)}
\label{Ssyml}
\end{equation}
which follow from (\ref{Sapprox}), together with the approximations (\ref{Gasym2}) to evaluate the weightings $\textup{f}_g^*(q)$ and $\textup{f}_l^*(q)$. Fig.~\ref{Fig5} demonstrates the excellent agreement between this analytical approximation and that obtained from the numerical solution of the OZ equation (\ref{OZS}) for the asymmetric potential (\ref{phiasym}). Again, as with the correlation function $G(0,0;q)$, the difference between both results can only be discerned by looking at the percentage error, as shown in the inset.\\

We also present in Fig.~\ref{Fig5} the approximation for $S(0;q)$ described in \cite{Parry2019}. This is derived, in an alternative manner, using the resonance expansion (\ref{Res}), which does not involve computing symmetric gas and symmetric liquid potentials, and is written
\begin{equation}\label{Snat}
\begin{split}
S(0;q)\;\approx\; \textup{f}_g(q)\,S_g(q)+\textup{f}_l(q)\,S_l(q)\;+\\[.25cm]
\frac{\,\Delta\rho\,\rho'(0)}{\sigma q^2}\;\left(\textup{f}_g(q)\,\frac{S_g(q)}{S_g(0)}+\textup{f}_l(q)\;\frac{S_l(q)}{S_l(0)}\right)
\end{split}
\end{equation}
This expression involves weights, different to those appearing in (\ref{Gasym1new}) and (\ref{Sasym}), given by $\textup{f}_g(q)=\Delta\kappa_g(q)/(\Delta\kappa_g(q)+\Delta\kappa_l(q))$ with $\Delta\kappa_b(q)=\sqrt{\kappa_b^2+q^2}-\kappa_b$, and $\textup{f}_g(q)+\textup{f}_l(q)=1$. As seen in Fig.~(\ref{Fig5}), it is remarkable to us that these two approximations, which are very different in appearance, are almost identical numerically, and each describe accurately the behaviour of the local structure factor over the whole wave-vector range. We comment on this later.\\

We can also determine the properties of the local structure factor away from the origin, since the solution of the OZ equation can be written exactly as
\begin{equation}\label{Sz}
S(z;q)=\left\{
\begin{array}{ll}
S_g^\textit{sym}(z;q)+A\;G_g^\textit{sym}(0,z;q) & \text{for}\;z<0\\[.25cm]
S_l^\textit{sym}(z;q)-A\;G_l^\textit{sym}(0,z;q) & \text{for}\;z>0
\end{array}\right.
\end{equation}
where
\begin{equation}
A\;=\;\frac{S_l^\textit{sym}(0;q)-S_g^\textit{sym}(0;q)}{G_g^\textit{sym}(0,0;q)+G_l^\textit{sym}(0,0;q)}
\end{equation}
In (\ref{Sz}), we may use the exact resonant expansion (\ref{Res}) for $S_b^\textit{sym}(z;q)$ on each side of the interface. As expected, there are two different sets of surface tension-like coefficients $\{\sigma_n^b\}$, one in the liquid side ($b=l$), one in the gas side ($b=g$). 
When the position $z$ is a few correlation lengths away from the origin, only the term related to the derivative of the profile is of importance, and we can approximate, at fixed $q$,
\begin{equation}
S(z;q)\;\approx\; S_b(q)\;+\;\frac{\Delta\rho\,\rho'(z)}{\,\sigma_1^b\, q^2(1+\xi_b^2q^2)\,}
\label{PRE}
\end{equation}
Here, the surface tension-like coefficients are only determined by bulk properties \cite{Parry2016}
\begin{equation}
\sigma_1^b\;=\;\kappa_b\,\Delta\rho\,\frac{\Delta \phi''(\rho_b)}{\,|\Delta\phi'''(\rho_b)|\,}
\label{sigma1}
\end{equation}
Finally, integration of $S(z;q)$ over the macroscopic interval $[-L_g,L_l]$ determines the total structure factor exactly as
\begin{equation}
\begin{split}
S_\textup{\tiny TOT}(q)=\frac{1}{2}\left(S_\textup{\tiny TOT g}^\textit{sym}(q)+S_\textup{\tiny TOT l}^\textit{sym}(q)\right)\\
-\frac{1}{2}\;
\frac{\left(S_l^\textit{sym}(0;q)-S_g^\textit{sym}(0;q)\right)^2}{G_g^\textit{sym}(0,0;q)+G_l^\textit{sym}(0,0;q)}
\end{split}
\end{equation}
This may then be evaluated to high accuracy by using the approximations (\ref{Gasym2}), (\ref{Ssymg}) and (\ref{Ssyml}) together with the approximations for the symmetric total structure factors \cite{Parry2019} 
\begin{equation}
S_\textup{\tiny TOT b}^\textit{sym}(q)\;\approx\; 2\,L_b\,S_b(q)\,+\,\frac{4(\rho_0-\rho_b)^2}{\sigma_b^\textit{sym}\,q^2}\,\frac{S_b(q)}{S_b(0)}
\end{equation}
It is straightforward to show that, with these approximations, the total structure factor $S_\textup{\tiny TOT}(q)$ contains the correct Goldstone mode divergent term $\Delta\rho^2/\sigma q^2$ in the limit $q\to 0$, which is not sensitive to capillary-wave fluctuations, since there is no dependence on the derivative of the profile.

\section{Sullivan model}

Having demonstrated that liquid-gas asymmetry can be treated within square-gradient, we now show that near identical results apply to the more microscopic Sullivan model for which we may use the very accurate Carnahan-Starling equation of state for hard-spheres. The Sullivan model is based on the Helmholtz free-energy functional \cite{Sullivan1979,Sullivan1981,Tarazona1982,Parry2016}
\begin{equation}
F[\rho]=\!\int\!\! d{\bf r}\,f_h\big(\rho({\bf r})\big)\,+\,\frac{1}{2}\int\!\!\!\int\! d{\bf r}_1d{\bf r}_2\;\rho ({\bf r}_1)\,
w(|{\bf r}_1-{\bf r}_2|)\, \rho ({\bf r}_2)
\label{300}
\end{equation}
where $f_h(\rho)$ is the free-energy density for bulk hard-spheres, and the  derivative $d f_h(\rho)/d\rho=\mu_h(\rho)$ is the local hard-sphere chemical potential. The intermolecular potential has a Yukawa form
\begin{equation}
w(r)\;=\;-\frac{\alpha}{\,4\pi rR^2\,}\;e^{-r/R}
\end{equation}
where $-\alpha\!=\!\int\! d{\bf r}\, w(r)$ is the integrated strength and the range is $R$. Hereafter, without loss of generality, we set $R=1$ or, equivalently, measure all lengths in units of $R$. Minimization of the grand potential functional leads to the Euler-Lagrange equation for the density profile which, for a planar interface, reduces to
\begin{equation}\label{ELsull}
\frac{d^2\mu_h}{dz^2}\;=\,\mu_h\big(\rho(z)\big)\,-\,\mu\,-\,\alpha\,\rho(z)
\end{equation}
which is also similar to the ODE for the profile in square-gradient theory. This has a first integral
\begin{equation}
\frac{d\mu_h}{dz}\;=\sqrt{\psi(\mu_h)}
\end{equation}
where
\begin{equation}
\psi(\mu_h)\;=\;\big(\mu_h(\rho)-\mu\big)^2-\,2\alpha \big(p_h(\rho)-p\big)
\end{equation}
and $p_h(\rho)$ is the hard-sphere pressure. The function $\psi(\mu_h)$ plays the same role as the bulk potential $2\Delta\phi(\rho)$ in square-gradient theory. It has equal minima at $\mu_h(\rho_b)$ and a maximum at $\mu_h(\rho_0)$, where $\rho_0$ is the density at which $d^2\mu_h/dz^2=0$. This we choose as the origin. The surface tension $\sigma$ can be written \cite{Sullivan1981}
\begin{equation}\label{SullTen}
\sigma\;=\;\frac{1}{\alpha}\int\! dz\, \left(\frac{d\mu_h}{dz}\right)^2,
\end{equation}
and is similar to the square-gradient result. 

The correlation functions and local structure factor also satisfy similar differential equations to those appearing square-gradient theory \cite{Parry2016,Parry2019}. If we define
\begin{equation}
\alpha\,H^{(2)}(z,z';q)\;=\; \mu'_h\big(\rho(z)\big)\,\mu'_h\big(\rho(z')\big)\; G(z,z';q)
\end{equation}
and  
\begin{equation}
  H(z;q)\;=\; \frac{S(z;q)\,\mu_h'\big(\rho(z)\big)}{1+q^2}\,,
  \label{H}
\end{equation}
manipulation of the OZ integral equation leads to the ordinary differential equations:
\begin{equation}
\left(-\partial^2_z +q^2+1-\alpha \frac{d\rho}{d\mu_h}\right)
H^{(2)}(z,z';q)\;=\,\delta(z-z')
\label{Greensull}
\end{equation}
and
\begin{equation}
\left(-\partial^2_z +q^2+1-\alpha \frac{d\rho}{d\mu_h}\right)
H(z;q)\;=\,1
\label{OZSsull2}
\end{equation}
The latter ODE identifies the bulk structure factors $S_b(q)$ in the liquid ($b=l$) and gas ($b=g$) phases as
\begin{equation}
S_b(q)=\frac{S_b(0)}{\;1+\displaystyle\frac{q^2\xi_b^2}{1+q^2R^2}\;}
\label{310}
\end{equation}
where
\begin{equation}
S_b(0)=\frac{1}{\;\mu_h'(\rho_b)-\alpha\;}
\end{equation}
and identifies the OZ correlation lengths $\xi_b$ as
\begin{equation}
\frac{R}{\xi_b}\;=\;\sqrt{\,\frac{\mu_h'(\rho_b)}{\alpha}-1\,}
\end{equation}
where we have reinstated the lengthscale $R$. Similarly, the 2D Fourier transform of the pair-correlation function in the bulk is given by 
\begin{equation}
G_b(q)\;=\;\frac{\alpha (d\rho/d\mu_h)^2}{2\sqrt{\kappa_b^2+ q^2}}
\end{equation}
where $\kappa_b=1/\xi_b^T$ is the inverse of the bulk true correlation length $\xi^T_b =\sqrt{\xi_b^2+R^2}$.

The functions $H(z;q)$ and $H^{(2)}(z,z';q)$ are related in precisely the same way that $S(z;q)$ and $G(z,z';q)$ are within square-gradient theory, provided that $d\rho/dz$ is replaced by $d\mu_h/dz$. It follows that, if we specialise to a class of free-energy densities $f_h(\rho)$ that have an Ising symmetry, we can write  $S(z;q)$ as an expansion over resonances \cite{Parry2019}:  
\begin{equation}\label{74}
\begin{split}
S(z;q)\;=\;S_b(q)\,\gamma_b(z)+\frac{\Delta\rho\,\rho'(z)}{\sigma_1\, q^2}\,\frac{S_b(q)}{S_b(0)}+\\
\frac{\Delta\rho}{\rho'(0)}\,\frac{S_b(q)}{S_b(0)}\,\sum_{n=2}^\infty\,\frac{\sigma}{\sigma_n}\,\frac{G\big(0,z;q\big)-G\big(0,z;\sqrt{n^2-1}\,\kappa_b\big)}{\big(1-\frac{q^2(\xi^T)^2}{n^2-1}\big)}
\end{split}
\end{equation}
where
\begin{equation}
\gamma_b(z)\;=\;\frac{\mu'_h(\rho_b)}{\mu'_h(\rho(z))}
\end{equation}
is referred to as the bulk-enhancement factor. Thus, apart from this additional factor $\gamma_b(z)$ in front of the first term, this is identical to the result for the square-gradient theory. Indeed, this allows us to immediately write the highly accurate approximations for the correlation function and local structure factor (at the origin) in symmetric systems as:
\begin{equation}
\frac{G(0,0;q)}{G_b(q)}\;\approx\;\, \gamma_b(0)^{\,2}\;+\;\frac{\rho'(0)^2}{\sigma q^2G_b(0)}
\label{GapproxSMsul}
\end{equation}
and
\begin{equation}
\frac{S(0;q)}{S_b(q)}\;\approx\; \gamma_b(0)\;+\;\frac{\Delta\rho\rho'(0)}{\sigma q^2S_b(0)}
\label{SapproxSMsul}
\end{equation}
An identical construction can now be used for a realistic hard-sphere chemical potential which will incorporate liquid-gas asymmetry. That is, symmetric gas and liquid systems can be defined as reflections of $\psi(\mu_h)$ about $\mu_h(\rho_0)$ where $\psi(\mu_h)$ is a maximum. The surface tension can then be written as the average (\ref{sigmaasym}),
where 
\begin{equation}
\sigma_g^\textit{sym}\,=\;\frac{2}{\alpha}\int_{-\infty}^0\!\!\! dz\; \left(\frac{d\mu_h}{dz}\right)^2
\end{equation}
and
\begin{equation}
\sigma_l^\textit{sym}\,=\;\frac{2}{\alpha}\int_0^\infty\!\!\! dz\; \left(\frac{d\mu_h}{dz}\right)^2
\end{equation}
Moreover, identical exact rules (\ref{Gasym1}) and (\ref{Sasym}) apply to the correlation function and local structure factor at the origin, defined where $d\mu_h/dz$ is a maximum (or, equivalently, when $\rho(z)=\rho_0$). 
These exact expressions can then be used in combination with the extremely accurate approximations
\begin{equation}
\frac{G_b^\textit{sym}(0,0;q)}{G_b(q)}\;\approx\;\, \gamma(0)^{\,2}\;+\;\frac{\rho'(0)^2}{\sigma_b^\textit{sym} q^2G_b(0;0)}
\label{Gsul}
\end{equation}
and
\begin{equation}
\frac{S(0;q)}{S_b(q)}\;\approx\; \gamma(0)\;+\;\frac{2|\rho_0-\rho_b|\,\rho'(0)}{\sigma_b^\textit{sym} q^2S_b(0)}
\label{Ssul}
\end{equation}
to determine the whole wave-vector dependence analytically. Thus, for example, the prediction for the correlation function at the origin incorporating liquid-gas asymmetry is
\begin{equation}\label{Gapprox2sul}
\frac{1}{G(0,0;q)}\;\approx\;\frac{1}{2}\,\sum_{b=g,l}\,\frac{1}{G_b(q)\left(\gamma_b(0)^2+
\frac{\rho'(0)^2}{\sigma_b^\textit{sym}\,q^2\,G_b(0)}\right)}
\end{equation} 

\begin{figure*}
\includegraphics[width=\columnwidth]{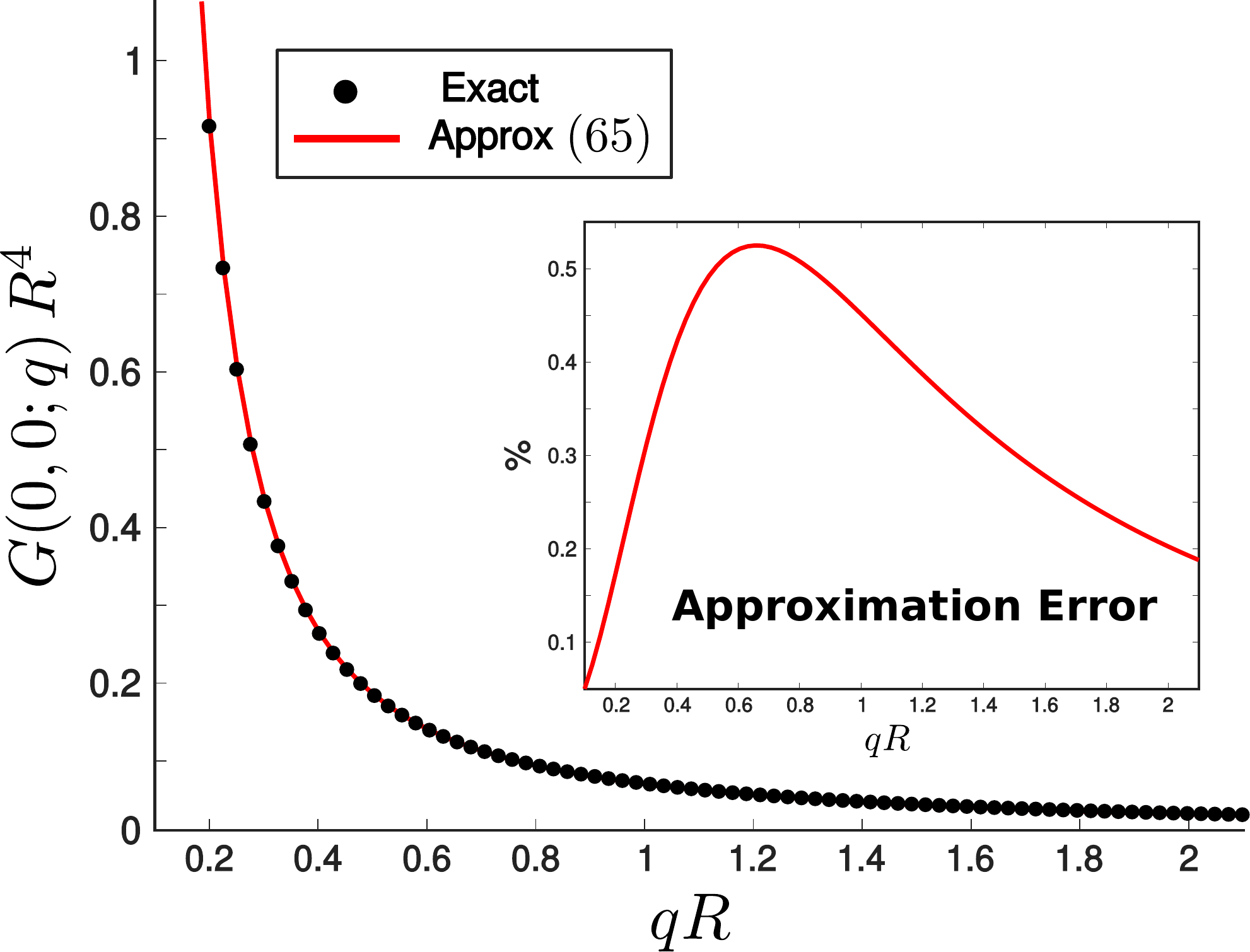}
\caption{\label{Fig6} Sullivan model results for $G(0,0;q)$. Comparison of the exact numerical solution (dots) of the OZ equation (\ref{Greensull}) with the analytical approximation (\ref{Gapprox2sul}). We use the Carnahan-Starling equation of state and a representative temperature $T/T_c=0.74$, corresponding to a strong asymmetry $\xi_l^T/\xi_g^T\approx 2.1$. The relative error of the approximation is shown in the inset.
}
\end{figure*}

In Fig.~\ref{Fig6}, we compare this approximation for $G(0.0;q)$ with the numerical solution of the OZ equation using the Carnahan-Starling equation of state for $\mu_h(\rho)$ for a representative temperature respresenting strong liquid-gas asymmetry. The maximum error is less that $0.5$ per cent. For the local structure factor, we also compare with the alternative approximation derived in \cite{Parry2019}
\begin{equation}\label{Ssul2}
\begin{split}
S(0;q)\;\approx\; \gamma_g(0)\,\textup{f}_g(q)\,S_g(q)+\gamma_l(0)\,\textup{f}_l(q)\,S_l(q)\\[.25cm]+
\frac{\,\Delta\rho\,\rho'(0)}{\sigma q^2}\;\left(\textup{f}_g(q)\,\frac{S_g(q)}{S_g(0)}+\textup{f}_l(q)\;\frac{S_l(q)}{S_l(0)}\right)
\end{split}
\end{equation}
where $\textup{f}_g(q)$ and $\textup{f}_l(q)$ are unchanged from (\ref{Snat}). Again, both approximations yield very similar results and are hardly distinguishable from the numerical solution of the OZ equation.

\section{Discussion}

In this paper, we have shown that, within both square-gradient theory and the more microscopic Sullivan model of the free interface, the presence of liquid-gas asymmetry can be elegantly incorporated using a simple construction which weights the correlation function and local structure factor in terms of the corresponding expressions for effective Ising symmetric liquid and gas interfaces. Using this approach, we can, for all practical purposes, analytically determine $G(0,0;q)$ and $S(0;q)$ over the full range of wave-vectors. Our approximations, in particular those for the correlation function (\ref{Gapprox2}) and (\ref{Gapprox2sul}), are near indistinguishable from the numerical solution of the OZ equation. These equations, for square-gradient theory and the Sullivan model respectively, are the main results of our paper.

To finish our paper, we make a few remarks regarding the interpretation of some of our results. The first of these is that the mixing rule for the correlation function (\ref{Gasym1}) is particularly simple. The reason why this mixing rule is expressed in terms of the inverse of the correlation function rather than $G$ itself can be traced to the Goldstone mode divergence, since the latter implies that $G$ is inversely proportional to the surface tension. There is, therefore, a simple correspondence between the mixing rule (\ref{Gasym1}) and the expression for the surface tension (\ref{sigmaasym}) written as the average of the symmetric gas and liquid tensions. It is interesting that, when expressed in terms of the weights $\textup{f}_g(q)$ and $\textup{f}_l(q)$ of the symmetric gas and symmetric liquid contributions, the exact mixing rules for the correlation function (\ref{Gasym1new}) and local structure factor (\ref{Sasym}) are the same. However, this parallelism does not imply that one can split consistently the correlation function and structure factor into bulk and interfacial contributions, as has been often assumed. In fact, our work shows that the correlation function and local structure factor decompose naturally into symmetric gas and symmetric liquid contributions, but not into bulk and interfacial contributions. This is clear in the explicit form of the approximation (\ref{Gapprox2sul}), which is for all practical purposes the analytical result for $G(0,0;q)$.

The mixing rule for the local structure factor leads to a new, accurate, analytical approximation for $S(0;q)$, which yields remarkably similar results to an alternate approximation (\ref{Ssul2}) proposed in \cite{Parry2019}. The reason for this robustness is  that both approximations rely on the constrained form of the structure factor imposed by the resonant expansion (\ref{Res}). The new approximation, based on (\ref{Sasym}) together with (\ref{Gsul}) and (\ref{Ssul}), is slightly more accurate than (\ref{Ssul2}), although the latter is easier to implement in simulations and experiments.

The analytical approximation (\ref{Gapprox2sul}) for the correlation function shows that liquid-gas asymmetry enters in three different ways: Through the different bulk contributions $G_b(q)$, through the surface tensions $\sigma_b^\textit{sym}$, which weight the Goldstone mode contributions, and finally through the bulk enhancements  $\gamma_b(0)^2$. In fact, the latter are the most important factors when the wave-vector $q$ is much larger than the inverse of the bulk gas correlation length, $q\gg\kappa_g$. In this limit, the correlation function and local structure factor behave as
\begin{equation}\label{Glim}
G(0,0;q)\;\to\;\; \frac{\alpha}{2\,\mu_h'(\rho_0)^2\,q}
\end{equation}
and
\begin{equation}\label{Slim}
S(0;q)\;\to\;\; \frac{1}{\mu_h'(\rho_0)}\left(1+\frac{1}{q^2}\right)
\end{equation}

Interestingly, these results are unrelated to the corresponding bulk liquid and gas expressions, and to the Goldstone mode behaviour. Indeed, they are entirely local, dependent only on the density $\rho_0=\rho(0)$, where $G$ and $S$ are being evaluated, as could be anticipated physically.  Also, we note that both expressions are equivalent to the large wave-vector behaviour of the correlation function and structure factor of the unstable, homogeneous, phase with density $\rho_0$. Note that the results (\ref{Glim}) and (\ref{Slim}) remain valid as the temperature is increased to the critical point, the interface disappears and $\rho_0$ tends to the critical density $\rho_c$. At this point, (\ref{Glim}) and (\ref{Slim}) correspond simply to the bulk critical behaviour of $G_b(q)$ and $S_b(q)$.

Finally, lets us return to the potential limitations of our method and conclusions. The models considered here are mean-field-like, and do not include the broadening of the interfacial width induced by capillary-wave fluctuations. However, as stressed earlier, in three dimensions the broadening of the profile is extremely weak and does not affect significantly the structure of the density profile, even for interfaces of mesoscopic extent. Indeed, we have already shown that the present approach quantitatively explains the complex wave-vector dependence of the total structure factor seen in the largest simulations of the liquid-gas interface in systems with truncated Lennard-Jones forces \cite{Hofling2015,Parry2016,Parry2019}. We are therefore confident that our predictions for the correlation function and local structure factor can be applied even in the presence of interfacial roughness, which would be included implicitly in the value of the gradient $\rho'(0)$. A second criticism is that our models are local and do not account for short-ranged packing effects. While packing effects do not significantly influence the density profile, they may well show up more prominently in the correlation function. However, it appears reasonable to us that the present local theories not only explain the leading-order  corrections to the Goldstone mode (as the wave-vector is increases) but identify correctly the physical meaning of the large-$q$ behaviour; namely, that the correlation function and structure factor behave locally. We believe that this is something that can be tested using DFT models that account more accurately for packing effects.

\acknowledgments

AOP acknowledges the EPSRC, UK for grant EP/L020564/1 (Multiscale analysis of complex interfacial phenomena). CR acknowledges the support of the grant PGC2018-096606-B-I00 (MCIU/AEI/FEDER, UE).

\bibliography{wetting}

\pagebreak

\begin{figure*}
\includegraphics[width=\columnwidth]{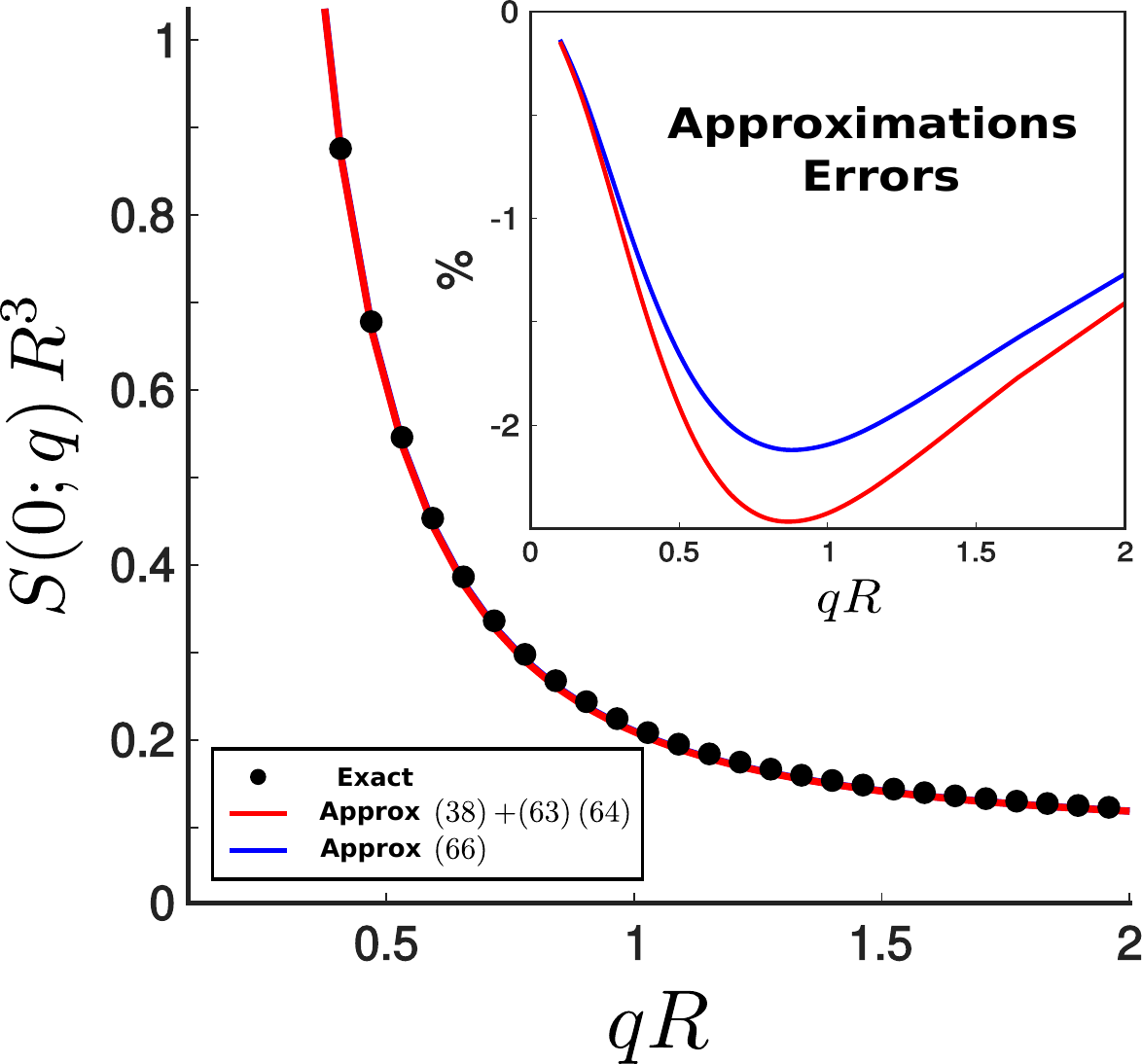}
\caption{\label{Fig7} Sullivan model results for $S(0;q)$. Comparison of the exact numerical solution (dots) of the OZ equation (\ref{OZSsull2}) for $T/T_c=0.74$ with the analytical approximation (\ref{Sasym}) together with (\ref{Gsul}) and (\ref{Ssul}) (red line). The alternative approximation based on (\ref{Ssul2}) (blue line) is also shown, but it is indistinguishable from the previous approximation. The relative errors of both approximations are shown in the inset.
}
\end{figure*}

\end{document}